\documentclass[prl,twocolumn,showpacs,amssymb,superscriptaddress,longbibliography]{revtex4-2}
\usepackage{graphicx}
\usepackage{amsmath}
\usepackage{amsfonts}
\usepackage{amsthm}
\usepackage{amssymb}
\usepackage{amsbsy}
\usepackage{wasysym}
\usepackage{bm}
\usepackage{bbm}
\usepackage{mathrsfs}
\usepackage{color}
\usepackage{hyperref}
\usepackage{braket}
\usepackage{soul}
\usepackage{soul}
\setcitestyle{super}

\newcommand{\defeq}{\mathrel{\mathop:}=}

\date{\today}
\begin{document}

\title{Microscopically exact transport equation for the quantum Calogero model}

\author{Vir B. Bulchandani}
\affiliation{Department of
Physics and Astronomy, Rice University, 6100 Main Street
Houston, TX 77005, USA}

\author{Fabian H. L. Essler}
\affiliation{Rudolf Peierls Centre for Theoretical Physics, University of Oxford, Oxford OX1 3PU, United Kingdom}

\begin{abstract}
We derive an exact transport equation for the quantum Calogero model, i.e. inverse-square interacting bosons on a line, from microscopic first principles.
\end{abstract}

\maketitle
\paragraph{Introduction.} In 1900, Hilbert posed his Sixth Problem of ``developing mathematically the limiting processes...which lead
from the atomistic view to the laws of motion of continua''~\cite{hilbert2000mathematical}. Much progress on Hilbert's Sixth Problem has been made in the classical setting~\cite{grad1958principles,cercignani1972boltzmann,lanford2005time}, culminating in the recent mathematical justification of the Boltzmann equation for classical hard spheres~\cite{deng2025hilbertssixthproblemderivation}. At the same time, the analogous problem for quantum systems has long resisted mathematical treatment~\cite{Erd_s_2004}, despite its fundamental importance for understanding transport phenomena in condensed matter physics~\cite{lifshitz2013statistical,kadanoff1962quantum,rammer2007quantum}. The quantum Boltzmann equation is instead usually justified heuristically by making two distinct approximations: the system is assumed to be in a weak-coupling or dilute regime, and correlation functions are assumed to factorize, in the sense that four-point and higher correlation functions are evaluated approximately using Wick's theorem~\cite{Erd_s_2004}.

A promising alternative arena for tackling Hilbert's Sixth Problem is the setting of classical or quantum integrable systems~\cite{faddeev1987hamiltonian,babelon2003introduction,korepin_quantum_1993,takahashi1999thermodynamics,sutherland2004beautiful,essler2005one,gaudin2014bethe}. Spatially local integrable systems exhibit an extensive number of commuting integrals of motion with spatially local densities, which results in the existence of stable quasiparticle excitations at all energy densities. This is in contrast to conventional, chaotic systems, for which quasiparticles at non-zero temperature generically acquire a finite lifetime.
% \fe{I think it would be good to add a short discussion of integrability here: existence of an extensive number of commuting integrals of motion with spatially local densities, which results in the existence of stable quasiparticle excitations at all energy densities.}
While integrable systems are therefore expected to be more mathematically tractable than microscopically chaotic examples such as hard spheres, the first-principles derivation of transport equations for such systems is nevertheless challenging. A notable breakthrough was the discovery of empirically exact transport equations --- so-called ``generalized hydrodynamics'' --- for integrable systems~\cite{Castro_Alvaredo_2016,bertini2016transport,doyon2020lecture,alba2021generalized,bouchoule2022generalized,essler2023short}. The power of generalized hydrodynamics stems from the fact that it does not invoke the two key approximations underlying the quantum Boltzmann equation~\cite{Erd_s_2004}, and is instead justified by passing to a ``ballistic'' or ``Euler'' scaling limit $x,t \to \infty$ with $x/t$ held constant~\cite{spohnlarge}. This is precisely the limit in which the Navier-Stokes equations recover their inviscid Euler limit.

Since the discovery of these equations in 2016, which can be made plausible at the physicist's level of rigour by analogy with classical soliton gases~\cite{percus1969exact,zakharov1971kinetic,boldrighini1983one,el2005kinetic,doyon2018soliton, bulchandani2018bethe}, much insight has been gained into their microscopic justification. At the same time, a first-principles derivation of the corresponding dynamical equation has remained elusive beyond classical~\cite{boldrighini1983one,el2005kinetic,croydon2021generalized} and non-interacting quantum~\cite{doyon2017large,Bastianello_2018,PhysRevB.96.220302,PhysRevX.12.041011} examples.  
% the most structured examples. \fe{What do you have in mind by the "most structured examples"?}

What do we mean by a first-principles derivation of generalized hydrodynamics? To proceed further, it is helpful to recall the form of the generalized hydrodynamics equations, which for continuous systems of quantum particles with a single flavour (the case studied in this work) take the form~\cite{Castro_Alvaredo_2016,bertini2016transport}
\begin{equation}
\label{eq:GHD}
\partial_t \rho(x,t,\lambda) + \partial_x (\rho(x,t,\lambda)v[\rho](x,t,\lambda)) = 0,
\end{equation}
where $\rho(x,t,\lambda)$ denotes the ``local density of states'' at position $x$, time $t$ and rapidity $\lambda \in \mathbb{R}$ (to be discussed further below) and $v[\rho](x,t,\lambda)$ is a model-dependent functional of the density of states $\{\rho(x,t,\lambda)\}_{\lambda \in \mathbb{R}}$ at each spacetime point, whose physical interpretation is the group velocity of the stable quasiparticle excitations, which exist as a result of integrability, in the macrostate specified by $\rho(x,t,\lambda)$. Eq.~\eqref{eq:GHD} is expected to hold in the Euler scaling limit. Crucially, it does not make the additional weak-coupling and Gaussian approximations underlying the quantum Boltzmann equation~\cite{Erd_s_2004}.

For the continuous particle systems of interest in this paper, the rapidity $\lambda$ recovers the bare momentum of each particle in the non-interacting limit of infinite interparticle separation~\cite{sutherland2004beautiful}. While one might expect an exact transport formalism for integrable systems to be expressed in terms of bare momenta, all available evidence suggests~\cite{spohn2023hydrodynamicscalesintegrablemanyparticle} that a description in terms of ``dressed momenta'' or rapidities, which only recover bare momenta in the limit of infinite interparticle separation, provides the correct extension of Boltzmann's approach for bridging the weakly and strongly coupled regimes of integrable gases. 
% \fe{I think the last two sentences (definition of the scaling limit) should be moved just below eqn (1).} 

The key property of $\rho(x,t,\lambda)$, which does not define it uniquely, is that in addition to satisfying Eq. \eqref{eq:GHD}, its moments should recover~\cite{Castro_Alvaredo_2016,bertini2016transport} local charge densities of the integrable system in question, i.e. $
\int_{-\infty}^{\infty} d\lambda \, \lambda^n \rho(x,\lambda, t) = q_n(x,t)$, where the global conserved charges of the model can be expressed as $\hat{Q}_n = \int dx \, \hat{q}_n(x,t,\lambda)$ and $q_n(x,t) := \mathrm{tr}[\hat{\rho} \hat{q}_n(x,t)]$ denotes an expectation value with respect to an arbitrary initial density matrix $\hat{\rho}$. Throughout this Letter, the operators $\hat{Q}_n$ will be taken to denote quantum conserved charges in involution, i.e. $[\hat{Q}_m,\hat{Q}_n] = 0$ for $m,n=0,1,\ldots$, with $\hat{Q}_2 \propto \hat{H}$ implying dynamical conservation of these charges.

In the classical setting, the derivation of Eq. \eqref{eq:GHD} is increasingly well understood. For several classical dynamical systems, including soliton gases~\cite{zakharov1971kinetic,el2005kinetic}, the hard-rod gas~\cite{percus1969exact,boldrighini1983one} and the ``box-ball system''~\cite{kuniba2020generalized,croydon2021generalized}, a first-principles microscopic derivation of Eq.~\eqref{eq:GHD} has been achieved. The more typical scenario is that there is no such derivation of Eq. \eqref{eq:GHD}, but the specific form of the quasiparticle velocity functional $v[\rho]$ can be derived microscopically in equilibrium~\cite{spohn2020collision,Bulchandani_2021,spohn2023hydrodynamicscalesintegrablemanyparticle,spohn2024currents,aggarwal2026asymptoticscatteringrelationtoda,aggarwal2026effectivevelocitiestodalattice,aggarwal2026fluctuations}. 
% This is the case for integrable systems that exhibit an extensive Lax matrix~\cite{spohn2023hydrodynamicscalesintegrablemanyparticle}, including the classical Toda~\cite{doyon2019generalized,spohn2020generalized,bulchandani2019kinetic} and Calogero models~\cite{Bulchandani_2021}. Such systems allow for an explicit identification of classical quasiparticles as Anderson-localized eigenvectors of the Lax matrix~\cite{bulchandani2019kinetic}, which enables a microscopic definition of the quasiparticle velocity $v[\rho]$ in equilibrium~\cite{bulchandani2019kinetic,Cao_2019,aggarwal2026asymptoticscatteringrelationtoda,aggarwal2026effectivevelocitiestodalattice,aggarwal2026fluctuations}.

The quantum setting is more challenging and to the best of our knowledge, there exists no first-principles microscopic derivation of Eq. \eqref{eq:GHD} in an interacting quantum system, i.e. one that is not unitarily equivalent to either free fermions~\cite{doyon2017large,Bastianello_2018,PhysRevB.96.220302} or free bosons~\cite{PhysRevX.12.041011}. We will elaborate further on some previous attempts below. However, as for classical systems, there exist microscopic methods for deriving the form of the quasiparticle velocity functional $v[\rho]$ in equilibrium states. Notable successes include the derivation of $v[\rho]$ from boost-type symmetries~\cite{Castro_Alvaredo_2016,yoshimura2020collision}, such as Galilean or Lorentz invariance and their lattice analogues and the exact treatment of current operators in spin chains~\cite{borsi2020current}.
% and the derivation of $v[\rho]$ using form-factor expansions~\cite{cubero2021form}. 
% \fe{Not sure what you have in mind here -- form factors are not needed for computing $v[\rho]$? The Cubero et al review also employ the "dressed form factor approach", which is based on strong assumptions that have never been verified. I am not fond of these works...}
Nevertheless, all these derivations currently fall well short of a first-principles derivation of the partial differential equation Eq. \eqref{eq:GHD}.

\paragraph{Summary of  results.} In this Letter, we present the first fully microscopic derivation of generalized hydrodynamics for an interacting quantum system, specifically the quantum (rational) Calogero model of inverse-square-interacting, bosonic quantum particles on a line, with first-quantized Hamiltonian
\begin{equation}
\label{eq:Hamiltonian}
\hat{H} = \sum_{i=1}^N \frac{\hat{p}_i^2}{2} + \sum_{i<j} \frac{\ell(\ell-\hbar)}{(\hat{x}_i-\hat{x}_j)^2}
\end{equation}
and interaction parameter $\ell \geq 0$. The Calogero model has attracted a great deal of attention since its discovery\cite{10.1063/1.1664820,10.1063/1.1664821,calogero1971solution}, with applications ranging from exotic statistics in quantum statistical mechanics \cite{PhysRevLett.67.937,bernard1994note,Murthy_1994,PhysRevLett.73.2150} to string theory \cite{jevicki1980semiclassical,polychronakos1992new,Polychronakos_2006} and mathematics \cite{dunkl1989differential,polychronakos1992exchange,etingof2007calogero}. A landmark result in the non-equilibrium physics of the Calogero model was the derivation of its zero-entropy-density fluid dynamics~\cite{Abanov_2005,Stone_2008,Abanov_2009,Abanov_2011}, which generalizes the Benjamin-Ono equation describing deep-water waves~\cite{benjamin1967internal,ono1975algebraic}. This description is limited to physics near zero-entropy states that allow for a description in terms of a Fermi surface at each spacetime point and is fully characterized by the local particle density and velocity fields. By contrast, generalized hydrodynamics is expected to hold at \emph{all} entropy densities, and allows for the initial specification of infinitely many independent, local charge densities $\{q_n(x,t=0)\}_{n=0}^{\infty}$ in the thermodynamic limit. Thus the previous results on the zero-entropy hydrodynamics of Calogero particles~\cite{Abanov_2005,Stone_2008,Abanov_2009,Abanov_2011} are a special limit of generalized hydrodynamics.

Away from the free boson ($\ell=0$) and free fermion ($\ell \to \hbar$) limits, the Calogero model is believed~\cite{polychronakos1989non} to be unitarily equivalent to a system of non-interacting particles with generalized (i.e. neither fermionic nor bosonic) statistics, which is consistent with its ideal-gas thermodynamics~\cite{bernard1994note,Murthy_1994,PhysRevLett.73.2150} and the lack of velocity dressing in its generalized hydrodynamics~\cite{Bulchandani_2021}. However, this unitary transformation does not admit an explicit description for more than two particles, even classically~\cite{polychronakos1989non}. Thus, even in the apparently simple setting of Calogero particles, a first-principles derivation of generalized hydrodynamics has proved elusive. 

We fill this lacuna by providing a fully microscopic definition of a local density of states $\rho(x,t,\lambda)$ that satisfies an exact transport equation 
\begin{equation}
\label{eq:CalogGHD}
\partial_t \rho(x,t,\lambda) + \lambda \partial_x \rho(x,t,\lambda) = 0
\end{equation}
expressed in terms of rapidities $\lambda$ rather than bare momenta. Our results go far beyond the usual justification for generalized hydrodynamics, Eq.~\eqref{eq:GHD}. In particular, we show that Eq.~\eqref{eq:CalogGHD} holds exactly even \emph{away} from the Euler scaling limit, while matching previous ballistic-scale predictions~\cite{Bulchandani_2021} from generalized hydrodynamics. Thus our results demonstrate that the effect of rational Calogero interactions is fully captured by the highly nonlinear mapping from momenta to rapidities, with no dispersive or collisional corrections to ballistic spreading.
% \fe{Shall we stress here that we do not have to take the Euler scaling limit?}
% Thus our results demonstrate that the effect of rational Calogero interactions is fully captured by the highly nonlinear mapping from momenta to rapidities, so that the ballistic-scale prediction from generalized hydrodynamics~\cite{Bulchandani_2021} is \emph{exact}, with no dispersive or collisional corrections to ballistic spreading.

We note that various first-principles approaches to deriving Eq. \eqref{eq:GHD} in quantum systems have been attempted in the past. One method involves perturbatively constructing an operator-valued Wigner function for interacting integrable systems, proceeding order-by-order in the interaction strength~\cite{bertini2022bogoliubov}. This matches the predictions of generalized hydrodynamics at leading order in the interaction parameter (in the present setting, this parameter corresponds to the deviation from the free-fermion point $\ell-\hbar$). By contrast, the method that we present here yields an equation of motion that is exact at all orders in $\ell-\hbar$. Another proposed approach is based on identifying the quasiparticles of quantum integrable systems with wavepackets, and then taking the semiclassical limit of the resulting gas of wavepackets~\cite{doyon2026towards}. While this motivates a ``phase-space density operator'' with a similar structure to the Wigner operator constructed here, it appears challenging to rigorously derive a transport equation for the former. In particular, we find that a version of Weyl ordering, rather than the ``classical'' or Kohn-Nirenberg~\cite{de2011preferred} ordering pursued previously~\onlinecite{doyon2026towards}, appears to be necessary to recover the generalized hydrodynamics equation Eq. \eqref{eq:GHD}, with Kohn-Nirenberg ordering yielding a distinct and nonlinear transport equation. 
% Finally, we note that exact hydrodynamic equations for the particle and momentum density have been proposed previously for the quantum Calogero model at zero temperature~\cite{Abanov_2005} but do not extend to the full density of states in Eq.~\eqref{eq:CalogGHD}.
% \fe{I wonder where the best place to discuss the works by Abanov et al is. We could have a discussion about known results of Calogero below (2), stressing its paradigmatic importance, and mention it there.}

Our main technical contribution is the construction of a Hermitian Wigner operator $\hat{W}(x,t,\lambda)$ that satisfies the following key properties:
\begin{enumerate}
    \item the Heisenberg-evolving operator $\hat{W}(x,t,\lambda)$ exactly satisfies the operatorial analogue of the free-streaming Boltzmann equation
    \begin{equation}
    \label{eq:operatorGHD}
    \partial_t \hat{W} + \lambda \partial_x \hat{W} = 0,
    \end{equation}
    \item the moments $\int_{-\infty}^{\infty} dx \int_{-\infty}^{\infty} d\lambda \, \lambda^n \hat{W}(x,t,\lambda) = \hat{Q}_n$ recover the exact quantum conserved charges $\hat{Q}_n$ of the quantum Calogero model,
    \item the position marginals $\int_{-\infty}^{\infty} d\lambda \, \hat{W}(x,t,\lambda) = \sum_{i=1}^N \delta(x-\hat{x}_i)$ recover the exact particle-density operator.
\end{enumerate}
Between them, Properties 1 and 2 ensure that the expectation value $\rho(x,t,\lambda) \defeq \mathrm{tr}[\hat{\rho}\hat{W}(x,t,\lambda)]$ with respect to any given initial density matrix $\hat{\rho}$ yields a non-perturbative, microscopic definition of the local density of states appearing in the generalized hydrodynamics equation Eq. \eqref{eq:GHD}. In particular, Property 1 guarantees that the classical expectation value $\rho(x,t,\lambda)$ satisfies Eq. \eqref{eq:CalogGHD}. Meanwhile, Property 2 guarantees that the moments $m_n(x,t) := \int_{-\infty}^{\infty} d\lambda \, \lambda^n \rho(x,t,\lambda)$ coincide with the expectation values of local charge densities $q_n(x,t)$ up to derivative terms, in the sense that their spatial integrals $\int_{-\infty}^{\infty} dx \, m_n(x,t) = \int_{-\infty}^{\infty} dx \, q_n(x,t) = \mathrm{tr}[\hat{\rho}\hat{Q}_n]$ recover the expectation values of the same conserved charges. 
% \fe{Shall we mention here, or somewhere else, that so far applications to experiments have mainly focused on the particle density and the position-integrated rapidity distribution, which can be respectively measured by time-of-flight experiments with 3D and 1D expansion? Our Wigner operator gives direct access to both, once the initial conditions have been specified.}
While Properties 1 and 2 seem to be essential features of any proposed derivation of generalized hydrodynamics for the quantum Calogero model, Property 3 could be relaxed in principle. However, it has the appealing consequence that the zeroth moment $m_0(x,t)$ coincides perfectly with the microscopic particle density. We note that so far, applications of generalized hydrodynamics to experiments have have mainly focused on the particle density and the position-integrated rapidity distribution, which can be measured by time-of-flight experiments with 3D and 1D expansion respectively~\cite{Schemmer_2019,malvania2021generalized,bouchoule2022generalized}. Our Wigner operator provides direct access to both these quantities once the initial conditions have been specified. We now proceed to the construction of $\hat{W}(x,t,\lambda)$ consistent with Properties 1-3 above.
\paragraph{Construction of the Wigner operator.} An unusual feature of the quantum Calogero model Eq. \eqref{eq:Hamiltonian} is that it admits a quantum Lax pair~\cite{wadati1993integrability,Bernard_1993,shastry1993super,Chalykh_2019}. Specifically, under Heisenberg-picture time evolution, which we henceforth assume, the $N$-by-$N$ operator-valued matrices $\hat{X}_{jk} = \hat{x}_j \delta_{jk}$
and $\hat{L}_{jk} = \hat{p}_j \delta_{jk} + \frac{i\ell}{\hat{x}_{j}-\hat{x}_k}(1-\delta_{jk})$ satisfy the relations
\begin{align}
\label{eq:Lax1}
\frac{d}{dt}\hat{X} + i[\hat{X},\hat{A}] &= \hat{L}, \\
\label{eq:Lax2}
\frac{d}{dt}\hat{L} + i[\hat{L},\hat{A}] &= 0,
\end{align}
where the operator-valued matrix $\hat{A}_{jk} = - \sum_{l\neq j} \frac{\ell}{(\hat{x}_{j}-\hat{x}_{l})^2} \delta_{jk} + \frac{\ell}{(\hat{x}_j-\hat{x}_{k})^2}(1-\delta_{jk})$. In contrast to the classical theory of Lax pairs, the quantities $\mathrm{tr}[\hat{L}^n]$ are not necessarily conserved under the quantum dynamics~\cite{wadati1993integrability}; a complete set of conserved charges is instead given by $\hat{Q}_n = \mathbf{v}^T \hat{L}^n \mathbf{v}$, where the $N$-component vector $\mathbf{v} := (1,1,\ldots,1)^T$ and conservation of $\hat{Q}_n$ follows from the identity $\hat{A}\mathbf{v} = \mathbf{0}$.

We now propose to define a ``quantum empirical density of states'' or Wigner operator in terms of a \textit{two-dimensional} delta function in position-rapidity phase space, namely
\begin{align}
\label{eq:defWigner}
\hat{W}(x,t,\lambda) = \mathbf{v}^T\delta^2(x-\hat{X}(t),\lambda-\hat{L}(t))\mathbf{v}.
\end{align}
This two-dimensional delta function can be defined explicitly as the limit $\delta^2(\hat{A},\hat{B}) : = \lim_{\epsilon \to 0} \frac{1}{\pi \epsilon} e^{-(\hat{A}^2+\hat{B}^2)/\epsilon}$. The latter definition enforces two-dimensionality of the delta function, which in turn enforces Weyl ordering of the operators $\hat{X}$ and $\hat{L}$. This specification would not be necessary for commuting operators $[\hat{A},\hat{B}]=0$, for which there is no natural distinction between a two-dimensional delta function and a product of one-dimensional delta functions; indeed, for commuting operators the factorization $\delta^{2}(\hat{A},\hat{B}) = \delta(\hat{A})\delta(\hat{B})$ is preserved under the Gaussian regularization above. Crucially, this property does \emph{not} hold when $[\hat{A},\hat{B}] \neq 0$; if we instead define the Kohn-Nirenberg ordered Wigner operator $\hat{W}_{\mathrm{KN}}(x,t,\lambda) = \mathbf{v}^T \delta(x-\hat{X}(t))\delta(\lambda-\hat{L}(t)) \mathbf{v}$, which is closer to definitions in previous work~\cite{Bulchandani_2021,doyon2026towards}, this yields a non-closed dynamical equation that differs from Eq. \eqref{eq:operatorGHD}.

In order to establish Properties 1-3 above, it will be simplest to work with the multivariate Taylor expansion of $\hat{W}(x,t,\lambda)$, namely (suppressing arguments)
\begin{equation}
\label{eq:seriesexpansion}
\hat{W} = \sum_{m,n=0}^\infty \frac{(-1)^{m+n}}{m!n!} \mathbf{v}^T \mathcal{S}[\hat{X}^m \hat{L}^n]\mathbf{v} \delta^{(m)}(x)\delta^{(n)}(\lambda).
\end{equation}
Here $\mathcal{S}$ denotes the ``Weyl ordering''~\cite{hall2013quantum} of $\hat{X}^m \hat{L}^n$, where Weyl ordering is understood to apply to the operator-valued matrices $\hat{X}$ and $\hat{L}$, rather than to the single-particle canonical operators $\hat{x}_j$ and $\hat{p}_j$ as is more conventional in physics. Thus $\mathcal{S}[\hat{X}^m\hat{L}^n]$ denotes the uniform average over all $C_{m,n} = \frac{(m+n)!}{m!n!}$ distinct orderings of the $m+n$ symbols making up the monomial $\hat{X}^m\hat{L}^n$, so that for example $\mathcal{S}[\hat{X}\hat{L}] = \frac{1}{2}(\hat{X}\hat{L}+\hat{L}\hat{X})$ and $\mathcal{S}[\hat{X}^2\hat{L}] = \frac{1}{3}(\hat{X}^2\hat{L} +\hat{X}\hat{L}\hat{X} + \hat{L}\hat{X}^2)$. In what follows, it will also be useful to note the ``filtering identity'' $\int_{-\infty}^\infty dy\, y^{m}\delta^{(n)}(y) = (-1)^n n!\delta_{mn}$ for integers $m,n \geq 0$.

We first establish Property 3: integrating over $\lambda$ in Eq. \eqref{eq:seriesexpansion}, using the filtering identity and resumming the resulting Taylor series yields $
\int d\lambda \, \hat{W}(x,t,\lambda) = \mathbf{v}^T\delta(x-\hat{X})\mathbf{v} = \sum_{j=1}^N \delta(x-\hat{x}_j)$ by definition of $\hat{X}$. We next establish Property 2: applying the filtering identity twice, we obtain $\int_{-\infty}^\infty dx \int_{-\infty}^{\infty} d\lambda \, \lambda^n \hat{W}(x,t,\lambda) = \mathbf{v}^T \hat{L}^n \mathbf{v} = \hat{Q}_n$, which non-trivially recovers the conserved charges of the quantum Calogero model. 

It remains to prove Property 1. To this end, define operators $\hat{O}_{m,n}(t) := \mathbf{v}^T \mathcal{S}[\hat{X}^m(t)\hat{L}^n(t)] \mathbf{v}$ for integers $m,n \geq 0$. It follows by Eqs.~\eqref{eq:Lax1} and \eqref{eq:Lax2}, the permutation invariance of Weyl ordering~\cite{hall2013quantum}, and the identity $A\mathbf{v} = \mathbf{0}$ and its transpose, that
\begin{equation}
\frac{d}{dt} \hat{O}_{m,n}(t) = \begin{cases} 0, & m=0, \\
m\hat{O}_{m-1,n+1}(t), & m \geq 1.
\end{cases}
\end{equation}
In Eq. \eqref{eq:seriesexpansion}, this implies that
\begin{equation}
\label{eq:dtW}
\partial_t \hat{W} = \sum_{m=1}^{\infty}\sum_{n=0}^{\infty} \frac{(-1)^{m+n}}{(m-1)!n!} \hat{O}_{m-1,n+1} \delta^{(m)}(x) \delta^{(n)}(\lambda).
\end{equation}
Meanwhile, the distributional identity
\begin{equation}
\lambda \delta^{(n)}(\lambda) = \begin{cases} 0, & n =0, \\ -n\delta^{(n-1)}(\lambda), & n\geq 1,\end{cases}
\end{equation}
which follows by integration over smooth test functions, implies that
\begin{equation}
\label{eq:dxW}
\lambda \partial_x \hat{W} = -\sum_{m=0}^{\infty}\sum_{n=1}^{\infty}\frac{(-1)^{m+n}}{m!(n-1)!}\hat{O}_{m,n}\delta^{(m+1)}(x)\delta^{(n-1)}(\lambda).
\end{equation}
The series in Eqs. \eqref{eq:dtW} and \eqref{eq:dxW} coincide upon relabelling indices and Eq.~\eqref{eq:operatorGHD} follows. Taking expectation values $\rho(x,t,\lambda) := \mathrm{tr}[\hat{\rho}\hat{W}(x,t,\lambda)]$, we deduce that $\rho(x,t,\lambda)$ provides the desired microscopic definition of the local density of states in Eq. \eqref{eq:CalogGHD}.

Finally, we note that the arguments above immediately yield a microscopic derivation of Eq.~\eqref{eq:CalogGHD} for the \emph{classical} Calogero model. This can be deduced either by passing to the semiclassical limit $\hbar \to 0$ with $\ell$ fixed in the above expressions, or by working purely classically, i.e. starting from the classical Calogero Hamiltonian 
\begin{equation}
\label{eq:classicalH}
H = \sum_{i=1}^N \frac{p_i^2}{2} + \sum_{i<j} \frac{\ell^2}{(x_i-x_j)^2}
\end{equation}
and the corresponding classical~\cite{Abanov_2011} position and Lax matrices $X$, $L$ and $A$, and defining a ``classical Wigner operator'' $W(x,t,\lambda) = \mathbf{v}^T \delta^2(x-X(t),\lambda-L(t))\mathbf{v}$ just as in Eq. \eqref{eq:defWigner}. The resulting object is inequivalent to the Wigner operator $\widetilde{W}(x,t,\lambda) = \mathrm{tr}[\delta(x-X(t))\delta(\lambda-L(t))]$ proposed for the classical Calogero model in previous work~\cite{Bulchandani_2021}. Indeed, while both the classical Wigner operators $W$ and $\widetilde{W}$ satisfy the classical versions of Properties 2 and 3 above, only $W$ satisfies the classical version of Property 1 that gives rise to Eq.~\eqref{eq:CalogGHD}.

\paragraph{Initial conditions and a consistency check.} In order to apply Eq.~\eqref{eq:GHD} in practice, we need to specify its initial conditions $\rho(x,0,\lambda)$. This requires converting an initial density matrix $\hat{\rho}$ into its corresponding distribution function $\rho(x,0,\lambda)$, a problem whose solution remains largely open beyond free-fermion limits~\cite{doyon2017large,Fagotti_2020}.
% \fe{Shall we also cite \onlinecite{doyon2017large} here? "has never been mathematically justified..."-> "a problem whose solution remains largely open beyond..."?}
% \fe{I think it would be helpful for non-experts to explicitly state that in order to apply (1) in practice, we need to know the initial conditions, which requires translating an initial density matrix $\hat\rho$ into its corresponding distribution function $\rho(x,0,\lambda)$, which is general problem for GHD.}
Instead, the generalized hydrodynamics equations Eq. \eqref{eq:GHD} are generally understood by physicists as pertaining to a somewhat loosely defined class of ``local equilibrium'' initial states, i.e. states that locally resemble a generalized Gibbs state~\cite{spohnlarge,Castro_Alvaredo_2016,bertini2016transport,Fagotti_2020}. For example, in modeling quantum experiments near thermal equilibrium it is natural~\cite{bouchoule2022generalized} to consider local-equilibrium initial density matrices of the form $\hat{\rho} \propto \exp\left(-\sum_{n=0}^{\infty} \int_{-\infty}^\infty dx\, \beta_n(x) \hat{q}_n(x)\right)$, with the local Lagrange multipliers $\beta_n(x)$ smooth and slowly varying functions of position $x$.

A striking conclusion of our analysis for the quantum and classical rational Calogero models is that for inverse-square interactions, no such local-equilibrium assumption is necessary: for our microscopic definition of $\rho(x,t,\lambda)$, Eq.~\eqref{eq:CalogGHD} holds \emph{regardless} of the choice of initial state. Nevertheless, a natural extension of our results, which we do not attempt here, would be an exact specification of the initial states $\rho(x,0,\lambda)$ corresponding to local-equilibrium initial density matrices. We expect that this will recover the local density approximation that is standard~\cite{Castro_Alvaredo_2016,bertini2016transport} in the literature at zeroth order in the derivatives of $\beta_n$. Higher derivative corrections to the local density approximation are known to arise e.g. from emergent conformal invariance~\cite{Sotiriadis_2008,Langmann_2017,doyon2017large,bulchandani2020hydrodynamics} at low temperatures, or from the interplay between non-zero scattering lengths and initial spatial inhomogeneity~\cite{,Bulchandani_2024,PhysRevLett.132.251602}. Based on Eq. \eqref{eq:CalogGHD}, we conjecture that the only corrections to the generalized hydrodynamics of the rational Calogero model are such corrections to the initial local density approximation. 

We can test this conjecture indirectly in the classical setting by considering an exact solution to the dynamics of the classical Calogero Hamiltonian Eq. \eqref{eq:classicalH}. In particular, it was shown~\cite{Abanov_2011} that the classical dynamics of Eq. \eqref{eq:classicalH} has the exact ``zero-soliton'' solution $x_j(t) = \sqrt{(\ell/\omega)}\sqrt{1+\omega^2 t^2}c_j$, where $\omega$ is a free parameter and $c_1 < c_2 < \ldots < c_N$ are the roots of the $N$th Hermite polynomial $H_N(c_j)=0$. For this solution, it is easily verified that the classical particle density $n(x,t) = \sum_{j=1}^N \delta(x-x_j(t))$ at time $t$ is related to the particle density at time $t=0$ by a simple rescaling $n(x,t) =n(x/\sqrt{1+\omega^2t^2},0)/\sqrt{1+\omega^2t^2}$. Moreover, in the large-$N$ limit, this density profile exhibits a semicircle law 
\begin{equation}
\label{eq:semicirclelaw}
n(x,t) \to \frac{N}{\pi R(t)^2/2}\sqrt{R(t)^2-x^2}, \quad N \to \infty,
\end{equation}
where $R(t)^2 := 2N(\ell/\omega)(1+\omega^2 t^2)$ and corrections to this expression are subleading in $1/N$~\cite{mehta2004random,Abanov_2011}. A highly non-trivial check on generalized hydrodynamics and the local density approximation is that they recover the exact time evolution Eq.~\eqref{eq:semicirclelaw}.

To model the initial state $x_j(0) = \sqrt{(\ell/\omega)}c_j$ in the local density approximation, we simultaneously make the large-$N$ approximation in Eq. \eqref{eq:semicirclelaw} and note that since $p_j(0) = 0$ for this initial condition, it can be regarded as being locally at zero temperature. Pursuing the resulting local density approximation, whereby the state at each position $x$ is assumed to be the unique zero-temperature Gibbs state with the initial density in Eq. \eqref{eq:semicirclelaw} (as obtained from the thermodynamic Bethe ansatz for Calogero models~\cite{bernard1994note,Bulchandani_2021}), yields the approximate initial condition $\rho_{\mathrm{LDA}}(x,0,\lambda) = \frac{1}{2\pi \ell}\Theta(\pi \ell n_{\mathrm{LDA}}(x,0) - |\lambda|)$, where $\Theta(x)$ denotes the Heaviside step function and we model the initial density profile as $n_{\mathrm{LDA}}(x,0) = \frac{N}{\pi R(0)^2/2}\sqrt{R(0)^2-x^2}$. Evolving this initial condition under Eq.~\eqref{eq:CalogGHD} yields the prediction $n_{\mathrm{LDA}}(x,t) = \int d\lambda \, \rho_{\mathrm{LDA}}(x-\lambda t, \lambda, 0) = \frac{N}{\pi R(t)^2/2}\sqrt{R(t)^2-x^2}$, in perfect agreement with the exact result Eq. \eqref{eq:semicirclelaw}. This leads to the remarkable conclusion that for this exact solution to the classical Calogero dynamics, the \emph{only} possible corrections to generalized hydrodynamics and the local density approximation are $1/N$-type corrections to the semicircle law Eq. \eqref{eq:semicirclelaw}, which is consistent with the claimed exactness of the dynamical equation Eq.~\eqref{eq:CalogGHD}.
\paragraph{Conclusion.} We have derived generalized hydrodynamics equations for the quantum and classical Calogero models from microsopic first principles. We have further shown that in conjunction with the local density approximation, these equations match a known exact solution \cite{Abanov_2011} to the microscopic classical dynamics at leading order in a large-$N$ expansion. The former result in particular represents, to the best of our knowledge, the only microscopically exact derivation of a transport equation for an interacting quantum Hamiltonian of bosons or fermions.

Our results raise several questions for future work. A simplifying feature of inverse-square interactions is the absence of velocity dressing in the thermodynamic Bethe ansatz~\cite{Bulchandani_2021}, and therefore of hydrodynamic diffusion in these systems~\cite{gopalakrishnan2018hydrodynamics,De_Nardis_2018}. Can the methods introduced in this paper be extended to settings with velocity dressing? A promising arena for exploring this extension of our results is the hyperbolic Calogero model, which exhibits velocity dressing in addition to being defined in the continuum and possessing a Lax pair~\cite{spohn2023hydrodynamicscalesintegrablemanyparticle}. A potential complication is that the Lax matrix generating the conserved charges of the quantum hyperbolic Calogero model~\cite{romer1996conservation} is distinct from the Lax matrix that recovers the rapidities of the classical hyperbolic Calogero model~\cite{spohn2023hydrodynamicscalesintegrablemanyparticle}. Another desirable extension of our results would be to the Haldane-Shastry spin chain, which lacks velocity dressing like the rational Calogero models~\cite{bulchandani2024hydrodynamics}. This would require a better understanding of the microscopic operators that create non-interacting spinons in the Haldane-Shastry chain, which has long proved elusive~\cite{talstra1997creation}, perhaps because there is no simple correspondence between Bethe quasiparticles and the spectrum of a quantum Lax matrix in this setting~\cite{inozemtsev1990connection,Bernard_1993,Talstra_1995}. Nevertheless, we expect that the Haldane-Shastry chain will be more tractable than the related~\cite{haldane1994physics}, experimentally relevant~\cite{lake2013multispinon,scheie2021detection,wei2022quantum,rosenberg2024dynamics} quantum Heisenberg chain.

More generally, we hope that the microscopic derivation of a transport equation presented in this paper will prove useful for benchmarking past and future proposals for deriving transport equations for interacting quantum systems from microscopic first principles. 

\paragraph{Acknowledgments.} V.B.B. thanks H. Ha and A.G. Abanov for helpful discussions on this topic. We thank the organizers and hosts of the Les Houches workshop on ``Novel Emergent Phenomena in Quantum Many-Body Dynamics" in August 2024, where this work was initiated. This work was performed in part at the Aspen Center for Physics, which is supported by National Science Foundation grant PHY-2210452.
\bibliography{bibl}
\end{document}